# Interferometric speckle visibility spectroscopy (iSVS) for measuring decorrelation time and dynamics of moving samples with enhanced signal-to-noise ratio and relaxed reference requirements

Yu Xi Huang[1], Simon Mahler[1], Jerome Mertz [2,3], Changhuei Yang,[1,*]

[1]*Department of Electrical Engineering, California Institute of Technology, Pasadena, California 91125, USA*
[2]*Department of Biomedical Engineering, Boston University, Boston, Massachusetts 02215, USA*
[3]*Neurophotonics Center, Boston University, Boston, Massachusetts 02215, US*

*\*chyang@caltech.edu*

**Abstract:** Diffusing wave spectroscopy (DWS) is a group of techniques used to measure the dynamics of a scattering medium in a non-invasive manner. DWS methods rely on detecting the speckle light field from the moving scattering media and measuring the speckle decorrelation time to quantify the scattering medium's dynamics. For DWS, the signal-to-noise (SNR) is determined by the ratio between measured decorrelation time to the standard error of the measurement. This SNR is often low in certain applications because of high noise variances and low signal intensity, especially in biological applications with restricted exposure and emission levels. To address this photon-limited signal-to-noise ratio problem, we investigated, theoretically and experimentally, the SNR of an interferometric speckle visibility spectroscopy (iSVS) compared to more traditional DWS methods. We found that iSVS can provide excellent SNR performance through its ability to overcome camera noise. We also proved iSVS system has more relaxed constraints on the reference beam properties than most other interferometric systems. For an iSVS to function properly, we simply require the reference beam to exhibit local temporal stability, while incident angle, reference phase, and intensity uniformity do not need to be constrained. This flexibility can potentially enable more unconventional iSVS implementation schemes.

## 1. Introduction

Diffusing wave spectroscopy (DWS) is a group of techniques that use light transmitted through a scattering medium to extract the dynamics and the properties of the scattering medium [1–7]. DWS methods rely on detecting the transmitted speckle light field from the scattering media and measuring the speckle decorrelation time as a quantifier of the scattering medium's dynamics. The chief advantage of DWS lies in its non-invasiveness when used to probe biological samples. The technique is widely used in biological studies, such as measuring cerebral blood flow through tissues and bones [8–10]. In DWS, the speckle decorrelation time between the measured data is extracted as a metric to provide the dynamic information of the scattering medium, and the speckle decorrelation time is typically calculated in two different methods: temporal sampling and spatial ensemble.

Temporal sampling based DWS is also called diffuse correlation spectroscopy (DCS) [8–16]. In general DCS applications, the laser light propagates through a dynamic scattering medium and the transmission is collected back by using a single photon avalanche diode (SPAD) or a photodetector. Typically, a single speckle or a small collection of speckles (constituting a single aggregate speckle) is detected at a sufficiently high frame rate such that the inverse of the frame rate is significantly shorter than the expected decorrelation time of the speckle [17–23]. The typical DCS algorithm then calculates the signal's time correlation to measure the scattering medium's dynamics. In biological applications, DCS generally employ a fast (typically >100 kHz sampling rate) sensor with high quantum efficiency and low detector noise characteristics, such as an ultrafast SPAD or TOF, to better detect the typically weak signals.

There are different and more advanced forms of DCS. In the interferometric DCS (iDCS) [18–23], a reference light field is introduced to enable interferometric detection of the optical signal. This provides the ability to reject stray light and enable interferometric suppression of the detector noise when measuring the optical signal. It is also possible to simultaneously detect multiple speckles with multiple high-speed detectors, such as a SPAD array [17–19,21]. Recently, a method that combines interferometry and ultrafast cameras for multiple speckle detection has also been reported that sought to combine both advantages to yield a stronger DCS signal [24].

The alternate DWS approach is based on the use of spatial ensemble. This class of method is generally referred to as the speckle visibility spectroscopy (SVS) [25–27]. Other approaches utilizing spatial ensembles of speckles include laser speckle contrast imaging (LSCI) [28–36]. Here, a high pixel count camera is used to simultaneously detect multiple speckles in the optical transmission. Generally, the camera exposure time is set to be substantially longer than the speckle decorrelation time (typically a magnitude or more). This results in multiple different speckle patterns summing up in a single camera frame within the exposure time and yielding a blurred or washed-out speckle pattern. The speckle decorrelation time can then be calculated from the degree of blurring, or more specifically, the contrast of the speckles over all the detected speckles in the whole frame. By the nature of its operation, SVS does not require high-speed detectors and can work with mature and commercially available video-rate sCMOS cameras. In addition, SVS can naturally collect more transmitted light as video-rate cameras generally have very high pixel counts – the photon count per pixel may still be low, but a higher aggregate signal is obtained by using more pixels. On the other hand, SVS must contend with the relatively high detector noise characteristics of these cameras. To mitigate this issue, an interferometric SVS (iSVS) system has recently been demonstrated and shown to work well at suppressing detector noise through the interferometric measurement of the optical signal.

For DWS methods in general (both DCS and SVS), the accuracy of the speckle decorrelation time is a key specification. Mathematically, this accuracy can be quantified as the ratio of the measured decorrelation time to the standard error of the measurement, and it can be interpreted as the signal-to-noise performance specification of DWS methods. While DCS and SVS employ different methods to determine the speckle decorrelation time, their SNR performance can ultimately be tied back to the number of independent observables (NIO) and the detected photon flux [26]. The NIO corresponds to the number of speckle decorrelation events measured for single speckle DCS methods. The NIO then scales proportionately with the speckle count if multiple speckles are simultaneously detected for multi-speckle DCS methods. For SVS methods, the NIO corresponds to the number of speckle grains observed per exposure. Reference [26] details the exact NIO relationships for both DCS and SVS methods. The reference also compared the application of both DCS and SVS methods for biological studies, and found that for blood flow measurements, it is generally easier for SVS method to achieve a higher NIO count and, therefore, better accuracy in decorrelation time measurement. In addition, SVS methods cannot quantify the exact shape of the time decorrelation function, while DCS methods are generally able to do so.

The mathematical treatment of SNR performances in Ref. [26] assumes that the camera/detector is noise-free or that the camera/detector noise variance is low enough to have little impact on the measurements. While such an assumption is reasonable for a broad range of DWS applications, it will generally fail for low-light situations where the low number of detected signal photons can easily be overwhelmed by the intrinsic camera or detector noise, especially in biological applications where the exposure and emission levels are restricted [37].

This study aims to expand the mathematical model to account for detector noise. We will focus our analysis on an iSVS system, as such a system is specifically designed for low-light application uses, where interferometry is employed to enhance the detection of the signal light in the presence of significant detector noise. This advantage of iSVS over SVS has been experimentally demonstrated in previous studies [25]. It was shown that human cerebral blood flow could be determined with a high sampling rate of 100 Hz, even when the number of collected signal photons per pixel was low (i.e., below the camera noise variance) [25]. That promising result motivates our current analysis.

In this paper, we mathematically analyze and experimentally verify the signal-to-noise ratio (SNR) gains of iSVS over SVS in the presence of camera noise. We also show that with reasonable prior knowledge of the average signal intensity, the SNR of iSVS will always be higher or on par than SVS with a uniform reference field. Previous work used a uniform plane wave reference beam with a specific angle (off-axis interferometry) to enable cropping in the Fourier domain to select the signal [25,26]. We show that such requirements can be relaxed, and the results are nearly independent of the reference beam's shape and angle, allowing one to use a non-uniform wave reference, on-axis interferometry setup with no angle between reference and source, or even an entirely spatially speckled reference to achieve a similar result. We also deployed standard calibration methods for iSVS to calibrate noise in experiments, which is necessary for the system to operate in a wider range of intensity. We demonstrated that iSVS has the advantage of retrieving more consistent decorrelation time, especially in low-noise situations. Finally, we conclude that the properties discovered justify and simplify iSVS application implementation.

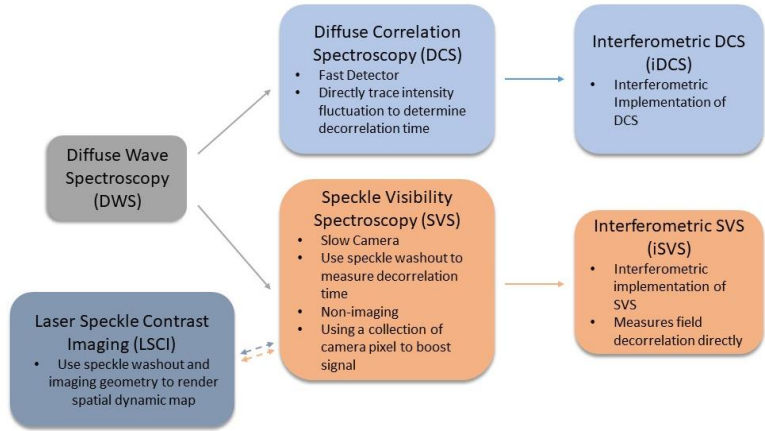

**Fig. 1.** Relationship and differences between DCS, SVS, and LSCI, as well as the interferometric methods iDCS and iSVS.

## 2. Theoretical derivations and results

In this section, we derive the SNR expressions associated with iSVS, then compare the theoretical performance of iSVS and SVS in the presence of camera noise. The SNR is defined by the ground truth decorrelation time $\tau$ and the standard error of the measured $\tau$ as:

$$SNR(\tau) = \frac{\tau}{SE(\tau)}, \qquad (1)$$

Using a stack of images, one can extract the speckle contrast for each image and determine $\tau$ in iSVS experiments. As such, we will first derive a mathematical expression for the speckle contrast as a function of $\tau$. We begin by assuming that the reference beam's intensity and phase are constant through time locally (i.e., at each pixel). We also assume that each camera pixel is the size of a single speckle grain, so each pixel observes an independent speckle. For details of the full derivation, please see the Supplemental Document. The iSVS intensity signal for pixel $j$ on the camera sensor, denoted as $I_j(t)$ is given by [25]:

$$I_j(t) = I_{Sj}(t) + I_{Rj}(t) + 2\sqrt{I_{Sj}(t)I_{Rj}(t)} \cos\left(\left(\phi_{Rj}(t) - \phi_{Sj}(t)\right)\right) + n_{shotj}(t), \quad (2)$$

which is composed of the sample intensity coming from the moving sample arm $I_{Sj}(t)$, the reference intensity from the reference arm $I_{Rj}(t)$, the interference cross term, and the shot noise $n_S(t)$. For notation, $\phi_{Sj}(t)$ is the phase of the sample beam, $\phi_{Rj}(t)$ is the phase of the reference beam, and any off-axis treatment or other adjustment to the reference is considered in the localized phase term $\phi_{Rj}(t)$. We note that in this derivation, we assume that the reference field can be spatially heterogeneous in both its amplitude, $I_{Rj}$, and phase, $\phi_{Rj}$, but that their values are fixed across all times. In effect, we have a heterogeneous and static reference field.

To obtain a reference-subtracted or reference-calibrated image, we subtract the captured image by the reference image. The reference-calibrated signal readout, in terms of photon counts, on pixel $j$ from the CMOS sensor can be mathematically expressed as:

$$N_j = \alpha \int_0^T I_j(t) - I_{Rj} \, dt = \alpha \int_0^T I_j^*(t) \, dt, \quad (3)$$

where $\alpha$ is the coefficient encompassing the quantum efficiency, the optical properties, and the other factors dictating the relationship between received photon counts and digitized readout signal. $I_j(t)$ is the intensity received by the pixel $j$ at time t, and $T$ is the exposure time. The speckle contrast $K$ of the entire image sensor is defined as [25,26]:

$$K = \frac{\sigma_{cam}}{\langle N_j \rangle} = \frac{\sqrt{\langle (N_j - \langle N_j \rangle)^2 \rangle}}{\langle N_j \rangle} = \frac{K_{up}}{K_{down}}. \quad (4)$$

As stated before, $N_j$ is the reference-calibrated photon count readout on pixel j, and the speckle contrast is the standard deviation divided by the mean. $\langle \cdot \rangle$ is the expected value of the term representing both an ensemble average and an average across all pixels.

$K_{up}^2$ can be expressed as:

$$K_{up}^2 = \left\langle \left(N_j - \langle N_j \rangle\right)^2 \right\rangle = \left\langle \left(\alpha \int_0^T \tilde{I}_j^*(t)dt\right)^2 \right\rangle, \quad (5)$$

Where $\tilde{I}_j^*(t)$ is the mean subtracted intensity on pixel $j$, or $\tilde{I}_j^*(t) = I_j^*(t) - \langle I_j^* \rangle$. We can then expand the integral as:

$$K_{up}^2 = \alpha^2 \left\langle \iint_0^T \tilde{I}_S(t_1)\tilde{I}_S(t_2) dt_1 dt_2 \right\rangle$$
$$+ 4\alpha^2 \left\langle I_{Rj} \iint_0^T \sqrt{I_{Sj}(t_1)I_{Sj}(t_2)} \cos\left(\Delta\phi_j(t_1)\right) \cos\left(\Delta\phi_j(t_2)\right) dt_1 dt_2 \right\rangle$$
$$+ \alpha^2 \left\langle \iint_0^T n_{shot_j}(t_1) n_{shot_j}(t_2) dt_1 dt_2 \right\rangle. \quad (6)$$

We can extract $I_{Rj}$ out from the second term as reference intensity is constant and independent from the sample intensity. Here, we model the scattering dynamics based on Brownian motion, leading to a field intensity autocorrelation function that decays exponentially with a decay constant depending on the decorrelation time $\tau_{field}$ [25,38,39]. In SVS, the intensity decorrelation time $\tau_s$ is first measured, and its relationship with the field decorrelation time $\tau_{field}$ can be extrapolated from the Siegert relation [6,40]. In iSVS, there are two decorrelation times influencing the speckle contrast: the first is related to the intensity of the sample arm ($\tau_s$), similar to SVS, and the second is related to the cross term from the interferometry which directly measures field decorrelation ($\tau_{field}$). Assuming the negative exponential decay holds, we approximate the decorrelation function as follows:

$$g_s(t = t_1 - t_2) = \frac{\langle \tilde{I}_S(t_1)\tilde{I}_S(t_2) \rangle}{\langle \tilde{I}_S(t_1)^2 \rangle} = e^{-\frac{t}{\tau_s}} = g_2(t) - 1 = |g_1(t)|^2, \quad (7a)$$

$$g_{cross}(t = t_1 - t_2) = \frac{\langle \sqrt{I_S(t_1)I_S(t_2)} \cos(\Delta\phi(t_1)) \cos(\Delta\phi(t_2)) \rangle}{\langle I_S(t_1) \cos^2(\Delta\phi(t_1)) \rangle} = e^{-\frac{t}{\tau_{field}}} = |g_1(t)|, \quad (7b)$$

$g_1(t)$ and $g_2(t)$ are the field and intensity decorrelation respectively, defined as $g_1(t) = \frac{\langle E(0)E^*(t) \rangle}{\langle I(0) \rangle}$, and $g_2(t) = \frac{\langle I(0)I(t) \rangle}{\langle I(0) \rangle^2} = 1 + |g_1(t)|^2$. Therefore, deriving from Eq. (7a) and Eq.

(7b), assuming the Siegert relation holds, $\tau_{field} = 2\tau_s$. Also, the shot noise is largely determined by reference intensity $\bar{I}_R$, and the Poisson statistics yield variance equal to $\alpha \bar{I}_R T$.

By substituting Eq. (7) into Eq. (6), we can obtain:

$$K_{up}^2 = \alpha^2 \langle \tilde{I}_S(t)^2 \rangle T \int_0^T 2\left(1 - \frac{t}{T}\right) g_s(t) dt$$
$$+ 4\alpha^2 \bar{I}_R \bar{I}_S T \langle \cos^2(\Delta\phi(t)) \rangle \int_0^T 2\left(1 - \frac{t}{T}\right) g_{cross}(t) dt + \alpha \bar{I}_R T, \qquad (8)$$

and through further simplification, with the assumption that $T \gg \tau_s$ and $T \gg \tau_{field}$, the integral $\left(1 - \frac{t}{T}\right) \approx 1$ before autocorrelation functions drop to 0. We obtain the variance of the calibrated image as:

$$K_{up}^2 \approx 2c\alpha^2 \bar{I}_S^2 T\tau_s + 4c\alpha^2 \bar{I}_S \bar{I}_R T\tau_{field} + \alpha \bar{I}_R T, \qquad (9)$$

where $c = 1$ (derived with the assumption that decorrelation follows exponential decay), but it can be generalized to gaussian or other forms, depending on the physics and the scattering dynamics.

$K_{down}^2$ can be expressed as the mean of the calibrated image, which only depends on the sample intensity as the reference is subtracted. By combining the denominator and numerator of the term $K^2$, we obtain its full relationship with $\tau_s$ and $\tau_{field}$:

$$K^2 \approx \frac{2\alpha^2 \bar{I}_S^2 T\tau_s + 4\alpha^2 \bar{I}_R \bar{I}_S T\tau_{field} + \alpha \bar{I}_R T}{\alpha^2 \bar{I}_S^2 T^2}. \qquad (10)$$

$\bar{I}_R$ is defined as the average intensity of the reference. $\bar{I}_S$ is defined as the average of the sample intensity. Both $\bar{I}_R$ and $\bar{I}_S$ represent the spatial and time average intensities across the camera. The variance, which is the numerator of $K^2$, has three main contributors: the variance from sample beam photons alone ($2\alpha^2 \bar{I}_S^2 T\tau_s$), the variance from the interference cross-term signal ($4\alpha^2 \bar{I}_R \bar{I}_S T\tau_{field}$), and the shot noise from the reference beam ($\alpha \bar{I}_R T$).

The above analysis assumes $I_{Rj}$ is much greater than $I_{Sj}$. For situations where the reference field is uniform, the condition is well satisfied for all pixels if it is well satisfied for any given pixel. When the reference field intensity is spatially heterogeneous, this condition may not be satisfied for all pixels. In this case, the above speckle contrast expression is still valid, except we would need to exclude pixels where $I_{Rj}$ is low compared to $I_{Sj}$ from our count. All averages and expectation values would only take into account the pixels where the condition $I_{Rj} \gg I_{Sj}$ is satisfied. In experiments, it is generally possible to arrange $\bar{I}_R$ to be much higher than $\bar{I}_S$, so that the $I_{Rj} \gg I_{Sj}$ is satisfied for the majority of the pixels. In such a case, the class of pixels to be excluded is so small that including them will only lead to a negligible error.

We preserved the first term in the numerator of Eq. (10) to show the contributions of both $\tau_s$ and $\tau_{field}$. In practice, we can further simplify the expression by noting that the first term is much smaller than the second and third numerator terms. If we set $U_I = \bar{I}_R/\bar{I}_S$, and define $N_{ST}$ as the average number of sample beam photons per camera pixel $N_{ST} = \alpha \bar{I}_S T$:

$$K^2 \approx U_I \left(\frac{4\tau_{field}}{T} + \frac{1}{N_{ST}}\right)$$
$$\approx U_I \left(\frac{8\tau_s}{T} + \frac{1}{N_{ST}}\right) \text{ (From Siegert Relation)}. \qquad (11)$$

The translation from $\tau_{field}$ to $\tau_s$ assumed Siegert relation holds. It is worth noting that the speckle contrast for SVS has the form of [26]:

$$K_{SVS}^2 = \frac{2\tau_s}{T} + \frac{1}{N_{ST}}. \qquad (12)$$

The forms are quite similar except for the different decorrelation time constants involved and the scaling term $U_I$.

The speckle contrast can be exactly determined for an ideal sensor with an infinite number of pixels so that the ensemble averages in the above calculation mirror the spatial averages. However, since only a finite number of pixels is available in a camera, the measured speckle contrast squared $\widehat{K^2}$ will contain an error, as $\widehat{K^2}$ will vary from measurement to measurement. The error will depend on the spatial profile of the reference beam and the NIO. We define a metric $R = \frac{\langle I_{Rj}^2 \rangle}{\langle I_{Rj} \rangle^2}$ to evaluate the uniformity of reference distribution. $R = 1$ represents a perfectly uniform reference intensity distribution, and a nonuniform reference pattern will have $R > 1$. As an example, a static speckled reference field where the speckle grain size is comparable or larger than the camera pixel will have an R-value approximately equal to 2. We also assume that the error of $K^2$ from measurement is only contributed by the numerator $K_{up}^2$ as $K_{down}^2 = N_{ST}^2$ can be accurately determined in experiments as a prior. Experimentally, we can determine $K_{down}^2 = N_{ST}^2$ exactly by blocking the reference beam and averaging arbitrarily many image frames.

This leads to a speckle contrast estimate given by:

$$\widehat{K^2} = K^2 \left(1 \pm \sqrt{\frac{3R-1}{NIO}}\right). \quad (13)$$

NIO is the number of independent observables as described in Ref. [26]. In our current analysis, NIO is equal to the number of camera pixels used in the iSVS and SVS experiment (recall that we are tackling a situation where the average speckle grain is equal to a camera pixel size). We use $\pm$ here to denote the standard deviation of the measurement-to-measurement variation.

Eq. (13) can be rewritten to provide a mathematical expression for determining $\tau_{field}$ and its measurement-to-measurement variation from the measured speckle contrast:

$$\hat{\tau}_{field} = \frac{T}{4U_I}K^2 - \frac{T}{4N_{ST}} \pm \frac{T}{4U_I}K^2\left(\sqrt{\frac{3R-1}{NIO}}\right). \quad (14)$$

This finally allows us to express the SNR of iSVS with either $\tau_{field}$ or $\tau_s$ from Siegert relation as (see Supplemental Document for full derivation):

$$SNR_{iSVS} = \frac{1}{\left(1 + \frac{T}{4\tau_{field}N_{ST}}\right)}\frac{\sqrt{NIO}}{\sqrt{3R-1}}. \quad (15)$$

From previous derivations in Ref [26], the SNR of SVS is:

$$SNR_{SVS} = \frac{1}{1 + \frac{T}{\tau_s N_{ST}}}\sqrt{\frac{NIO}{2}}. \quad (16)$$

We next consider the situation where camera noise is present in our analysis. We can define the camera noise term as a variable with mean $\langle n_{cam} \rangle = n_{offset}$ equal to the offset of the camera and variance $\sigma_{n_{cam}}^2$ related to the camera readout noise. We assume that the read noise is independent of the input signal strength $N_j$, independent of the exposure time $T$ of the camera and contains similar distribution on every pixel. The offset component will be eliminated when subtracting with the reference image; therefore, only the fluctuating variance survives. We define $\sigma_{n_{cam}}^2$ as the variance of the camera noise. Then by adding the variance contributed from the camera noise to $K_{up}^2$, we can rewrite $K^2$ for iSVS as:

$$K^2 \approx \frac{4\alpha^2 \overline{I_R I_S} T \hat{\tau}_{field} + \alpha \overline{I_R} T + \sigma_{n_{cam}}^2}{\alpha^2 \overline{I_S}^2 T^2}. \quad (17)$$

Then the $SNR_{iSVS}^{cam}$ is:

$$SNR_{iSVS}^{cam}(\tau_{field}) = \frac{1}{\left(1 + \frac{T}{4\tau_{field}}\left(\frac{1}{N_{ST}} + \frac{\sigma_{n_{cam}}^2}{U_I N_{ST}^2}\right)\right)}\sqrt{\frac{NIO}{3R-1}}. \quad (18a)$$

$$SNR_{ISVS}^{cam}(\tau_s) = \frac{\tau_s}{SE(\tau_s)} = \frac{\frac{\tau_{field}}{2}}{\frac{SE(\tau_{field})}{2}}$$

$$= \frac{1}{\left(1+\frac{T}{8\tau_s}\left(\frac{1}{N_{ST}}+\frac{\sigma_{n_{cam}}^2}{U_I N_{ST}^2}\right)\right)}\sqrt{\frac{NIO}{3R-1}} \quad (From\ Siegert\ Relation). \qquad (18b)$$

Similar to iSVS, for SVS, camera noise can be added to yield:

$$K^2 \approx \frac{2\alpha^2 \overline{I_S}^2 T\hat{\tau}_s + \alpha \overline{I_S} T + \sigma_{n_{cam}}^2}{\alpha^2 \overline{I_S}^2 T^2}. \qquad (19)$$

and:

$$SNR_{SVS}^{cam}(\tau_s) = \frac{1}{\left(1+\frac{T}{\tau_s}\left(\frac{1}{N_{ST}}+\frac{\sigma_{n_{cam}}^2}{N_{ST}^2}\right)\right)}\sqrt{\frac{NIO}{2}}. \qquad (20)$$

Consequently, in iSVS, when $N_{ST} \leq \sigma_{n_{cam}}$ (low signal case), the reference laser light field can boost the SNR (Eq. (18)) and substantially decrease the influence of camera noise because $U_I \gg 1$. When $N_{ST} \gg \sigma_{n_{cam}}$ (high signal case), the camera noise term can be ignored. In this case, the two system's SNR performance converge. Note the decrease of SNR from camera noise is due to the fluctuation of the camera noise variance; in the Supplemental Document we detailed how such fluctuation is eventually expressed with camera noise variance.

The above analysis indicates that, even though iSVS is an interferometry method, the requirements on the reference field are significantly relaxed. While most interferometry methods require the reference field to be planar and have a spatially invariant phase (or a phase with a linear spatial ramp), iSVS can be performed with a reference field with a spatially varying phase and amplitude. The SNR of iSVS is not penalized by a spatially varying phase and is only mildly penalized by a factor of $\sqrt{\frac{2}{3R-1}}$ for spatial heterogeneity. It should be noted that the number of independent observables (NIO) may vary under extreme heterogeneous reference cases, such as when some pixels have a reference intensity lower than the camera noise or the signal intensity, which breaks the assumption overlying the above analysis. In such cases, one should just count those pixels out and modify the NIO appropriately. However, these pixels are generally an overwhelming small class when the average reference intensity is set much higher than the sample intensity. In the following iSVS experiments, the reference intensity is sufficiently high such that less than 0.1% of the pixels had a reference intensity lower than camera noise and corresponding $\overline{I_S}$, and therefore, the effect of scattered reference illumination on NIO was negligible in the experimental data presented in this paper.

Figure 2 shows the SNR as a function of the decorrelation time for different averaged sample beam intensities $\overline{I_S}$, plotted with the following parameters: $\sigma_{n_{cam}}^2 = 8$ ph/pxl, $I_R \approx 3000$ ph/(pxl.ms), $NIO = 2000$, and $T = 200\ \mu s$. Camera noise variance is converted to keep consistency with the measured gray scale value. Note that the decorrelation time on the x-axis is the intensity decorrelation time. Also, the intensity units are in grayscale, but they should be comparable to later simulations and experiments. As evident, iSVS performs better than SVS at low signal intensities $\overline{I_S}$, but both uniform iSVS and SVS SNRs converge to the same performance as $\overline{I_S}$ is high enough to overwhelm the camera noise variance. The non-uniform reference penalizes the SNR, thus showing a lower SNR, but the advantage at low sample intensity levels still holds.

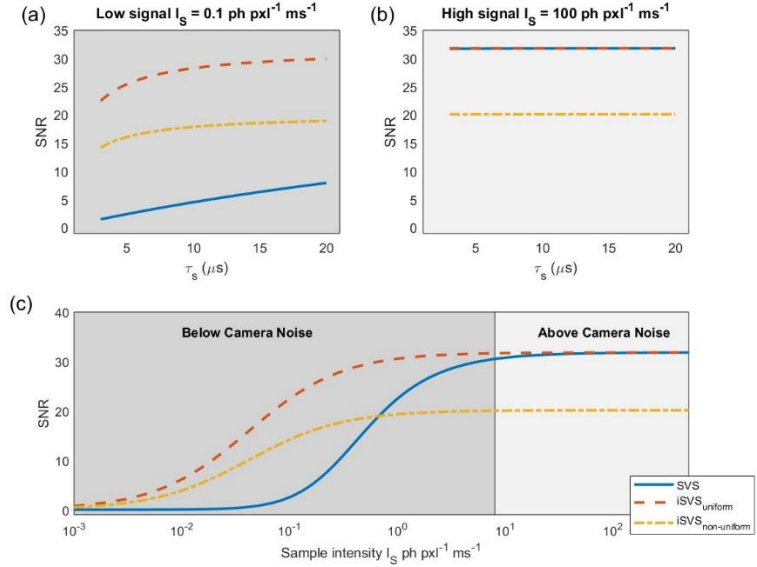

**Fig. 2.** Theoretical comparison between iSVS and SVS by using the SNR and the derived speckle contrast equations. (a), (b) SNR as a function of the decorrelation time at different averaged intensities $\overline{I_S}$ of the sample laser light field. Decorrelation time is the intensity decorrelation time $\tau_s$. (a) $\overline{I_S} = 0.1$ ph/(pxl.ms), (b) $\overline{I_S} = 100$ ph/(pxl.ms), (c) average SNR as a function of $\overline{I_S}$ for $\tau_s = [3, 20]\ \mu s$. As evident, iSVS performs better than SVS at low intensities $\overline{I_S}$, but both converge to same SNR as $\overline{I_S}$ overcome the camera noise variance. Parameters: $\sigma_{n_{cam}}^2 = 8$ ph/pxl, $I_R = 3000$ ph/(pxl.ms), $NIO = 2000$, and $T = 300\ \mu s$. Reference beam is uniform ($R = 1$) or speckled in iSVS non-uniform ($R = 2$).

## 3. Numerical results

Next, we performed numerical simulation by a designed speckle generation method based on speckle correlation [41]. We generated a series of speckle patterns with a chosen intensity decorrelation time. Then, the generated speckle patterns were combined into one camera image (direct imaging in SVS or interferometry in iSVS), and the speckle contrast was calculated. Detailed numerical simulation methods and other supporting simulation results can be found in the Supplemental Document.

The numerical results are presented in Fig. 3, showing the SNR as a function of the decorrelation time at different intensities $\overline{I_S}$ of the sample laser light field. In addition, iSVS with a randomly scattered reference laser light field ($R \approx 2$) was simulated. From Eq. (18), we expect that this arrangement will yield a SNR that is 30% worse than that for the uniform reference iSVS. Note that there is no specific unit for the generated intensity as well as time, since SNR is unitless, and therefore as long as the time and intensity units are consistent, the SNR result can be compared to experiment data.

Figure 3's results show that iSVS achieves a higher SNR at a low signal level of the sample laser light field, especially when the signal level is buried under camera noise variance. When the signal rises above the camera noise variance, the results from the two methods start to converge. The result suggests that iSVS can improve the SNR and counter the negative effect from the camera noise. Note the x-axis decorrelation time is still based on $\tau_s$. The slight fall-off of iSVS SNR at extremely large intensity is due to the breakdown of the assumption that $U_I \gg 1$.

We can observe in Fig. 3 that a speckled reference field iSVS case (heterogeneous reference field) has a slightly lower SNR than the uniform reference field iSVS case. We also note that the effective NIO is nearly identical between the uniform reference field and the speckled reference field; thus, the main performance difference is due to the heterogeneity of the speckle reference impacting the R factor.

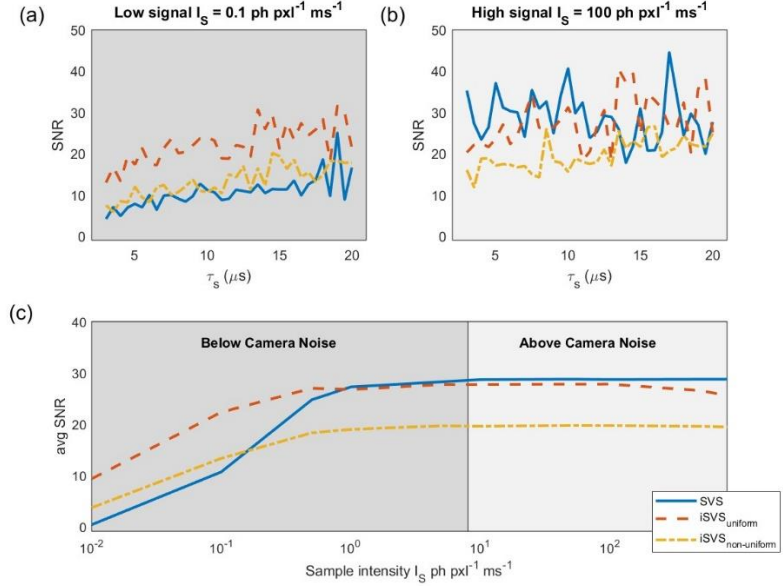

**Fig. 3.** Numerical simulation comparison between iSVS and SVS by using the SNR and the simulated speckle fields. (a), (b) SNR as a function of the decorrelation time at different averaged intensities $\bar{I}_S$ of the sample laser light field. (a) $\bar{I}_S = 0.1$ ph/(pxl.ms) (b) $\bar{I}_S = 100$ ph/(pxl.ms). The adjusted speckled reference curve is plotted by compensating the performance of speckled reference iSVS by the theoretical lost derived from Eq. (18). As evident, iSVS performs better than SVS at low intensities $\bar{I}_S$, but both converge to same SNR as $\bar{I}_S$ overcomes the camera noise variance (as in Fig. 2). Parameters: $\sigma_{n_{cam}}^2 = 8$, $\bar{I}_R = 3000$ ph/(pxl.ms), $NIO = 2000$, and $T = 200$ $\mu s$.

## 4. Experimental arrangement and results

To validate the theoretical and simulation results shown previously in sections 2 and 3, we designed an experiment to measure the SNR utilizing a rotating diffuser and an interferometer. The rotating diffuser is controlled by a DC motor, whose rotation speed is digitally controlled; thus, the decorrelation time of the speckle images is controlled electronically. The experimental arrangement is shown in Fig. 4. It is based on a Mach-Zehnder interferometer, whose input is a highly coherent (single-mode, single-frequency) continuous wave 523 nm collimated laser beam [Spectra-Physics Excelsior-532-200-LDRH]. Note that similar results should be obtained for any wavelength as long as the camera's quantum efficiency is reasonable at the illumination wavelength.

The input laser beam is first divided by a beam splitter ('BS1' in Fig. 4) into two light beams; one is the sample laser beam, and the other is the reference laser beam. The sample laser light beam propagates through a rotating diffuser and a static diffuser. The static diffuser counteracts the one-dimensional smearing effects from a single rotating diffuser, generating random speckles at the output. The reference laser light beam propagates in free space onto a second beam splitter ('BS2' in Fig. 4). A diffuser can be inserted in the path of the reference beam to generate a speckled reference laser light beam. We use tunable neutral density (ND) filters to control or to block (shut down) the intensities of the sample and the reference laser light beams. At the second beam splitter ('Beam Splitter 2' in Fig. 4), the two laser light beams interfere and form an interference pattern that is imaged to and recorded by a scientific CMOS camera sensor [PCO Edge 5.5] with an exposure time $T$.

To satisfy the prior of measuring $\langle N_j \rangle$ accurately, the reference laser light beam was blocked (at $t = 0$), creating a setup similar to an SVS setup, and $\langle N_j \rangle$ was estimated accurately from an average over 1000 images. This measurement is utilized later in calculating speckle contrast $K$ and $K^2$ for iSVS. Camera noise and its variance are also measured with both shutters closing, measuring only the camera background fluctuations.

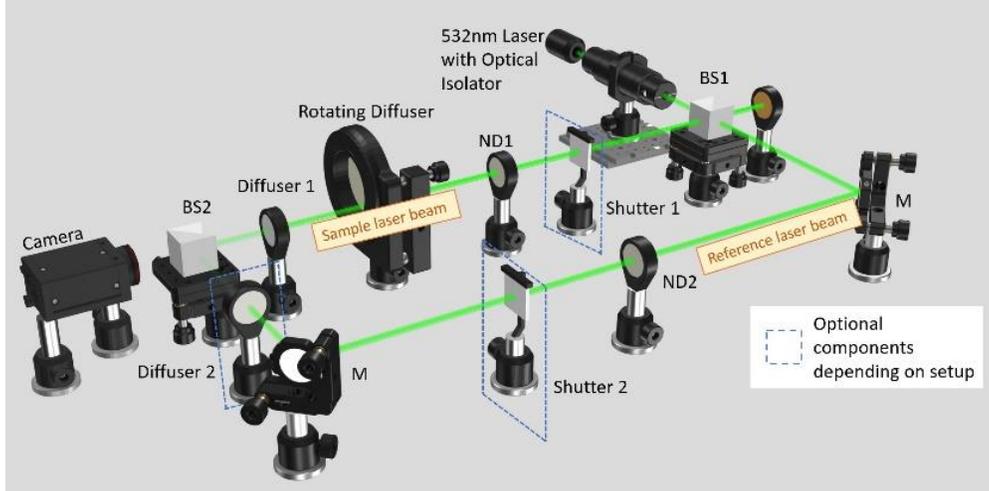

**Fig. 4.** Experimental arrangement for measuring the decorrelation time and SNR of a rotating diffuser, acting as moving sample. A diffuser 2 can be inserted to generate a distorted (randomly scattered) reference laser light beam in iSVS. In SVS, the reference laser beam is blocked.

A benchmark process is established to compare the performance of iSVS and SVS. The benchmark includes rotating the diffuser at 16 different rotational velocities. The rotation speed starts at 0 percent, then goes from 25 percent to 95 percent in 5 percent increments. The speeds are specifically chosen such that $\tau_s \ll T$ holds true. A set of reference $\tau_s$ is obtained by DCS experiment operating at a high signal level using an adjustable optical receiver [Newport 2051-FS] sampling at 1.25MHz. Then, the temporal autocorrelation curves are fitted with a negative exponential function, and decorrelation time is extracted for each rotation speed. Based on this benchmarking DCS reference experiment, the decorrelation time associated with the rotating diffuser was related to the rotational velocity of the DC motor, see the Supplemental Document.

For each rotational velocity, 200 images are captured with a frame rate of 150 FPS and exposure time $T = 300 \, \mu s$. For each of the 200 captured images, speckle contrast is calculated as shown in Fig. 1 for both iSVS and SVS. During SVS experiments, the reference laser beam path is blocked. First, SNR is calculated without considering the accuracy according to Eq. (1). The results are shown in Fig. 5.

Furthermore, despite incorporating the framework of shot noise and camera noise in theory, there are other external factors during experiments that would impact the noise profiles. To mitigate this and calibrate noise accurately, noise is measured instead of calculated: by closing both shutters, camera noise can be retrieved by calculating the variance of multiple measurements. Reference noise and sample arm noise can also be measured by keeping the diffuser in the static position and making multiple measurements. A total of 100 calibration measurements are taken for each trial during experiments, and noise is measured and factored in for both SVS and iSVS. Details of the calibration are presented in Supplemental Document.

We also observed additional noise terms with low spatial frequency characteristics in the experiment. These terms are likely attributable to the experimental spatial uniformity of the reference field and unaccounted laser mode instability. They introduce systematic errors into the experimental determination of $\tau_{field}$ and $\tau_s$. We addressed this by subtracting the empirically determined systematic error from the measured speckle contrast K prior to calculating the SNR. For the details regarding the offset calibration as well as results without calibration, please refer to the Supplemental Document.

We also calculated the $R$ values for the experimental reference beams. The distorted reference field pattern is speckled, but over 99.9% of the pixels are illuminated with a reference intensity larger than the camera noise variance, and therefore the effective NIO of the uniform reference iSVS and speckled reference iSVS can be considered equal. The uniform reference has a Gaussian spatial distribution with $R \approx 1.5$, while the speckled reference has $R \approx 1.9$. Therefore, the theoretical performance difference is about 15 percent—a relatively minor difference. Given the minor impacts of both considerations, we do not expect the two profiles of the reference beams used to significantly impact the SNR.

The SNR results are shown in Fig. 5. The ground truth is obtained by using the high sample intensity data. As detailed in the Supplemental Document, a series of noise calibrations is executed to subtract noise measured empirically. As evident in the data, the experimental results shown in Fig. 5 coincide with the theoretical and numerical results in Figs. 2 and 3. The overall trends in Fig. 5 match well with the expectation. iSVS performs better than SVS at low intensities $\overline{I_S}$, but both converge to the same SNR as $\overline{I_S}$ overcomes the camera noise variance, indicating no disadvantage present for performing interferometry. We note that given the similar NIO and other parameters, the SNR performance is slightly lower than theoretical and simulation values. This can be attributed to instability in the DC rotating motor speed at a high rotation speed. Laser noise in the form of mode hopping, laser intensity fluctuations, and/or environmental noise can also disrupt the coherence condition required for iSVS's interferometry scheme. These factors lead to variations in the measurements.

Figure 5(c) shows the SNR as a function of the average signal intensities $\overline{I_S}$ for the four cases: SVS, plane-wave uniform reference iSVS, and non-uniform speckled reference iSVS. Note that plane-wave reference iSVS and speckled reference iSVS exhibit an SNR index gain of at least two to three orders of magnitude higher at low signal situations than SVS. This allows the detection of the dynamics of the sample even if the signal level is below the camera noise. To achieve a similar SNR, iSVS and speckled iSVS both require 100 times fewer photons than SVS. Observe that there is no statistically significant performance difference between the speckled reference iSVS and the uniform reference iSVS in the experiment. The reason is likely because laser noise generates temporal fluctuations in reference intensity that equally affected both reference fields, overwhelming the 15% performance difference that may be presented with an ideal stable laser.

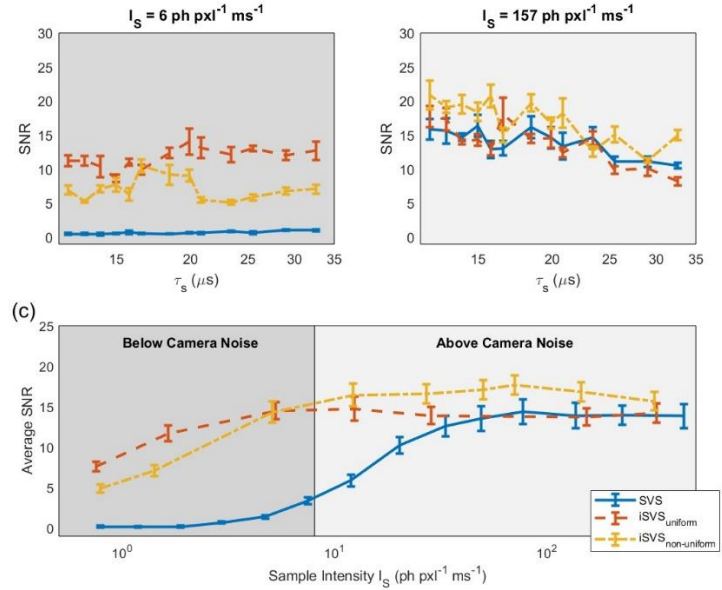

**Fig. 5** iSVS and SVS experiment results using the $SNR_{accuracy}$ defined in Eq. 21 and the measured speckle fields in the controlled experiment. (a), (b) SNR as a function of the decorrelation time at different averaged intensities $\bar{I}_S$ of the sample laser light beam. (a) $\bar{I}_S = 6$ ph/(pxl.ms), (b) $\bar{I}_S = 365$ ph/(pxl.ms). (c) Average SNR at various $\bar{I}_S$. As evident, iSVS performs better than SVS at low intensities $\bar{I}_S$, but both converge to same SNR as $\bar{I}_S$ overcome the camera noise variance (as in Fig. 2). Parameters: $\sigma_{n_{cam}}^2 = 8$ ph/pxl (converted from electron per pixel according to camera QE and conversion factor), $\bar{I}_R \approx 3000$ ph/(pxl.ms), $NIO \sim 2000$, and $T = 300\ \mu$s. X-axis $\tau_s$ derived DCS benchmark measurement associated with different rotation speeds.

## 5. Concluding remarks

In this study, we found that, by the merit of interferometry's ability to suppress camera noise, iSVS can accurately measure field decorrelation time in situations where the signal would otherwise be overwhelmed by camera noise in an SVS setup. This is particularly important in biomedical applications where the signal intensity tends to be limited and may otherwise require expensive cooled cameras or low-noise detectors to detect DWS signals. We also found that iSVS and SVS performed similarly when the signal intensity is high. This indicates that SVS can be expected to be sufficient in high sample intensity applications. Also, iSVS is more robust to other noises and can be easily calibrated to obtain consistent decorrelation times predictions, while SVS requires a larger signal to overcome the noise. In this study, we additionally found that the reference field used in iSVS can be spatially heterogeneous (unlike most other interferometry applications), with the heterogeneity only leading to slight performance degradation. This flexibility can potentially open the design space in iSVS system design. For instance, specific interferometer designs of the optical systems can achieve simultaneous direct measurement of both field and intensity properties, as shown in the last sections.

There is a potential advantage of iSVS that is not explored in this study – the ability to exploit temporal coherence to do time-gating. This capability can potentially be used in biomedical applications to select a sub-group of photon trajectories for detection. We anticipate that this is an area worth exploring in a follow-up study. Further investigation can also explore more in depth on different types of decorrelations and how it impacted the results in SVS. There are also potential to apply existing calibrations to further suppress the noise, either by measuring them more accurately or through filtering, increasing accuracy at various signal ranges.

In summary, we have demonstrated a unified approach to characterize the accuracy and SNR of iSVS method compared to SVS. As a prior publication has made a similar comparison between SVS and DCS methods, albeit without the presence of noise, we now have a unified analytical approach for comparing the major DWS methods.

**Funding.** This research was supported by the National Institutes of Health — Award No. 5R21EY033086-02. The research was also supported by Rockley Photonics, Inc. — Award No. Yang-Rockley-1.

**Acknowledgments.** The authors would also thank Kate Bechtel for helpful discussions and suggestions.

**Disclosures.** The authors declare no conflicts of interest.

**Data availability.** Data underlying the results presented in this paper are not publicly available at this time but may be obtained from the authors upon reasonable request.

**Supplemental Document.** See attached Supplemental Document for supporting content.

**Article thumbnail**

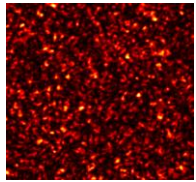